\documentclass[submission, Phys]{SciPost}
\usepackage{graphicx}
\usepackage{amsmath,amssymb}
\usepackage{subfig}
\usepackage{bm}
\usepackage{hyperref}
\usepackage{tikz} 

\usepackage{tikz,fp}
\usepackage{tikz-cd}
\usetikzlibrary{arrows}
\usetikzlibrary{intersections}
\usetikzlibrary{shapes.geometric}
\usetikzlibrary{decorations.pathmorphing, patterns,shapes,fixedpointarithmetic}
\usetikzlibrary{decorations.markings}

\pgfdeclarepatternformonly{south west lines}{\pgfqpoint{-0pt}{-0pt}}{\pgfqpoint{3pt}{3pt}}{\pgfqpoint{3pt}{3pt}}{
	\pgfsetlinewidth{0.4pt}
	\pgfpathmoveto{\pgfqpoint{0pt}{0pt}}
	\pgfpathlineto{\pgfqpoint{3pt}{3pt}}
	\pgfpathmoveto{\pgfqpoint{2.8pt}{-.2pt}}
	\pgfpathlineto{\pgfqpoint{3.2pt}{.2pt}}
	\pgfpathmoveto{\pgfqpoint{-.2pt}{2.8pt}}
	\pgfpathlineto{\pgfqpoint{.2pt}{3.2pt}}
	\pgfusepath{stroke}}

\tikzset{
	mid arrow/.style={postaction={decorate,decoration={
				markings,
				mark=at position .575 with {\arrow{stealth}}
	}}},
	near arrow/.style={postaction={decorate,decoration={
				markings,
				mark=at position .275 with {\arrow{stealth}}
	}}},
	far arrow/.style={postaction={decorate,decoration={
				markings,
				mark=at position .800 with {\arrow{stealth}}
	}}},
	snake arrow/.style={fixed point arithmetic, decorate, decoration={snake,amplitude=2pt, segment length=11pt},postaction={decoration={markings,mark=at position 0.625 with {\arrow{stealth}}},decorate}},
}

\newcommand{\RN}[1]{%
	\textup{\uppercase\expandafter{\romannumeral#1}}%
}
\newcommand{\Rn}[1]{%
	\textup{\lowercase\expandafter{\romannumeral#1}}%
}

\newcommand{\iu}{{i\mkern1mu}}
\newcommand*\diff{\mathop{}\!\mathrm{d}}

\hypersetup{
    colorlinks,
    linkcolor={red!50!black},
    citecolor={blue!50!black},
    urlcolor={blue!80!black}
}

\usepackage[bitstream-charter]{mathdesign}
\DeclareSymbolFont{usualmathcal}{OMS}{cmsy}{m}{n}
\DeclareSymbolFontAlphabet{\mathcal}{usualmathcal}

\begin{document}

	


        \begin{center}{\Large \textbf{
            Attenuating Dynamics of Strongly Interacting Fermionic Superfluids in SYK Solvable Models\\
            }}\end{center}
	\begin{center}
        Tian-Gang Zhou\textsuperscript{1} and
        Pengfei Zhang\textsuperscript{2$\star$}
        \end{center}

        \begin{center}
        {\bf 1} Institute for Advanced Study, Tsinghua University, Beijing,100084, China
        \\
        {\bf 2} Department of Physics, Fudan University, Shanghai, 200438, China
        \\
        ${}^\star$ {\small \sf pengfeizhang.physics@gmail.com}
        \end{center}

        \begin{center}
            \today
        \end{center}
	
	\section*{Abstract}
	{\bf 

        Quench dynamics of fermionic superfluids are an active topic both experimentally and theoretically. Using the BCS theory, such non-equilibrium problems can be reduced to nearly independent spin dynamics, only with a time-dependent mean-field pairing term. This results in persisting oscillations of the pairing strength in certain parameter regimes. However, experiments have observed that the oscillations decay rapidly when the interaction becomes strong, such as in the unitary Fermi gas \cite{KohlDecayRevivalTransient2021}. Theoretical analysis on this matter is still absent. In this work, we construct an SYK-like model to analyze the effect of strong interactions in a one-dimensional BCS system. We employ the large-$N$ approximation and a Green's function-based technique to solve the equilibrium problem and quench dynamics. Our findings reveal that a strong SYK interaction suppresses the pairing order. Additionally, we verify that the system quickly thermalizes with SYK interactions, whether it involves intrinsic pairing order or proximity effect, resulting in a rapid decay of the oscillation strength. The decay rates exhibit different scaling laws against SYK interaction, which can be understood in terms of the Boltzmann equation. This work represents a first step towards understanding the attenuating dynamics of strongly interacting fermionic superfluids.
         }
	
        \vspace{10pt}
        \noindent\rule{\textwidth}{1pt}
        \tableofcontents\thispagestyle{fancy}
        \noindent\rule{\textwidth}{1pt}
        \vspace{10pt}
	\section{Introduction}
	Non-equilibrium dynamics in systems with strong interactions is one of the most intriguing subjects in the realm of condensed matter and ultracold atoms. In particular, there is a growing interest in comprehending the quench dynamics, which involves observing the evolution resulting from a rapid change in parameters. Several seminal works from decades ago have explored novel quench dynamics in superconductors\cite{volkovCollisionlessRelaxationEnergy,AndersonRandomPhaseApproximationTheory1958,DzeroDynamicalVanishingOrder2006,FosterQuantumQuenchPhase2015,GurarieNonequilibriumDynamicsWeakly2009,LevitovSynchronizationBCSPairing2006,VarmaAmplitudeCollectiveModes1982,Tomadin2008}. Different dynamical phases are classified when adjusting the initial and final strength of attractive interactions according to the behavior of the pairing strength. In the phase diagram, one can observe the presence of persisting oscillations in the order parameter. This occurrence can be explained by the fact that the pioneering BCS theory can also be interpreted using the language of Anderson spins, where the oscillation corresponds to a collective mode of the Anderson spins within the mean magnetic field.
	
	However, the Fermi superfluids realized in ultracold gases may not be in a collisionless regime \cite{Balseiro2019} if a magnetic field is used to tune the scattering length between atoms \cite{KohlDecayRevivalTransient2021}. In particular, the unitary fermi gas is a typical strongly interacting system that can be realized in the experiment \cite{zwerger2011bcs,MukaiyamaMeasurementUniversalThermodynamic2010,StringariTheoryUltracoldAtomic2008,ThomasUniversalQuantumViscosity2011,ValeContactSumRules2019,ZwierleinRevealingSuperfluidLambda2012,ZwierleinSpectralResponseContact2019}. Unfortunately, the theoretical treatments of quench dynamics in unitary fermi gas don't reach a consensus yet. For a simple trial, we consider adding extra interaction between Anderson spins in addition to the BCS type mean-field interaction. For simplicity, we treat these additional interactions as all-to-all and Gaussian random interactions, inspired by the famous exact solvable Sachdev-Ye-Kitaev (SYK) model \cite{YeGaplessSpinfluidGround1993,kitaev2015simple,StanfordRemarksSachdevYeKitaevModel2016,YangConformalSymmetryIts2016,KamenevSachdevYeKitaev2016,Kitaev:2017awl,gu2017local,ZhangTunableQuantumChaos2017,ZhangCompetitionChaoticNonchaotic2017,XuInstabilityNonFermiliquidState2017,BalentsStronglyCorrelatedMetal2017,YaoQuantumCriticalityDuality2018,SachdevCriticalStrangeMetal2018,ZhaiTopologicalSachdevYeKitaevModel2018,KamenevQuantumCriticalityGranular2019,TarnopolskyNotesComplexSachdevYeKitaev2019,SachdevTransportChaosLattice2019a,ZhaoSymmetryBreakingCoupled2019,TarnopolskyNotesComplexSachdevYeKitaev2020,Chowdhury:2021qpy}. We assume the interaction is intra-spin and is independent for different spin componenets, which is different from models for the eternal traversable wormholes \cite{ZhaiDisconnectingTraversableWormhole2021,QiEternalTraversableWormhole2018,VerbaarschotQuantumChaosTransition2019,ZhangEntanglementEntropyTwo2019a,FranzRevivalDynamicsTraversable2020,ZhangCoupledSYKModel2020,FranzTraversableWormholeHawkingPage2020,ZhangTunnelingEternalTraversable2020,JafferisTraversableWormholeTeleportation2021,FranzTraversableWormholeCoupled2021,MilekhinSYKWormholeFormation2021,ZhangMoreComplexSachdevYeKitaev2021,Garcia-GarciaDominanceReplicaOffDiagonal2022}.
	
	We analyze the effect of SYK-type interactions in a one-dimensional BCS system, employing the large-$N$ approximation and a Green's function-based technique to investigate both the equilibrium problem and quench dynamics. Firstly, we examine the superconductivity transition point by calculating the critical hypersurface of the parameters. Subsequently, we explore the equilibrium phase diagram of pairing, considering both finite order parameters and BCS interaction. These phase diagrams consistently demonstrate that SYK interaction weakens the superconductivity. Furthermore, we numerically investigate the non-equilibrium quench dynamics, observing that the oscillation amplitude is suppressed by the SYK interaction $J$, which aligns with the findings from the equilibrium phase analysis. Finally, we observe that the decay rate exhibits distinct scaling laws with respect to the interaction $J$, depending on whether the pairing arises intrinsically or through the proximity effect. We argue that this behavior can be comprehended within the framework of the Boltzmann equation \cite{kamenev_2011}.

    \section{Model}
	\begin{figure}
		\centering
		\includegraphics[width=0.8\linewidth]{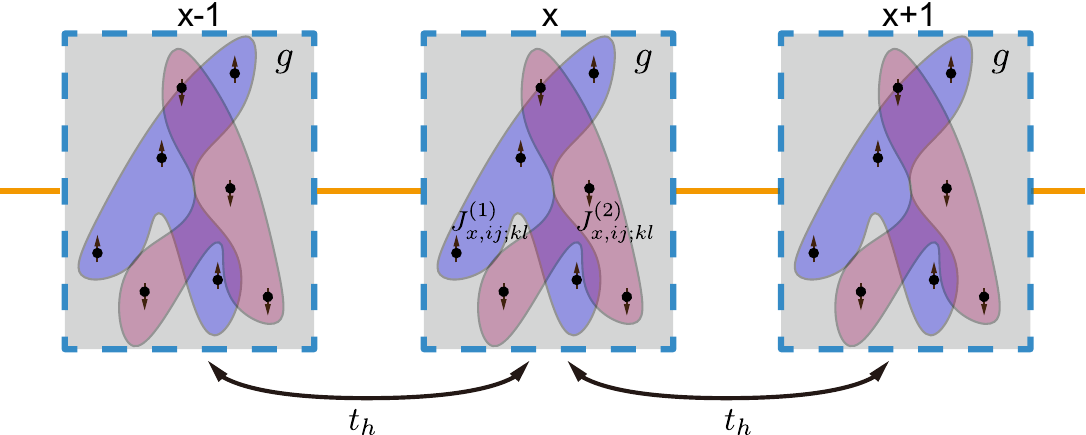}
		\caption{A pictorial representation of the model described by Eq.~\eqref{eq:H}, \eqref{eq:Hbcs} and \eqref{eq:HI}.
			Different dots with arrows represent different spin $1/2$ fermion in each unit cell. The blue and red blobs
			represent the intracell random interactions among the same spin and the solid lines with arrows denote
			the constant intercell hopping in the lattice. In this large-$N$ limit, the onsite attraction can be treated in the mean-field level, defined with interaction strength $g>0$.}
		\label{fig:illustration}
	\end{figure}

	Here we study the model in a one-dimensional spin-$1/2$ fermionic lattice model with up and down spin $l=\uparrow, \downarrow$. Depicted in Fig.~\ref{fig:illustration}, the Hamiltonian is composed of two parts: the one-dimensional BCS Hamiltonian and the intracell complex SYK-type interaction without any correlation between coupling constants for fermions with different spins. The Hamiltonian reads
	\begin{equation}\label{eq:H}
	\begin{split}
	\hat{H}= \hat{H}_{\text{BCS}}
			+ \hat{H}_{\text{I}},
	\end{split}
	\end{equation}
    The BCS Hamiltonian is 
    \begin{equation}\label{eq:Hbcs}
    	\begin{split}
    		\hat{H}_{\text{BCS}} =& \sum_{p,i,l}(\epsilon_p - \mu)\hat{c}^{\dagger}_{p,i,l} \hat{c}_{p,i,l}  
    		-\sum_{p,i}(\Delta \hat{c}^{\dagger}_{p,i, \uparrow} \hat{c}^{\dagger}_{-p, i, \downarrow} + \Delta^* \hat{c}_{-p, i, \downarrow} \hat{c}_{p, i, \uparrow} ),
    	\end{split}
    \end{equation}
    where $i=1,2,...,N$ labels different modes on a single site for each spin flavor. As a 1-d model, we assume the band dispersion as $\epsilon_p = - 2 t_h \cos(p)$, where $t_h$ is the hopping strength, and $p \in (-\pi, \pi]$.  $\mu = 0$ corresponds to half-filling because of the particle-hole symmetry after the disorder average. To proceed with equilibrium calculations, we need to distinguish two possible origins of order parameters $\Delta$. The $\text{type \RN{1}}$ comes from the background quantum proximity effect with a fixed order parameter, and the $\text{type \RN{2}}$ considers the self-consistent of the order parameter in evolution, which is
        \begin{equation}\label{eq:selfCcond}
        	\Delta = g/N_s \sum_p \langle \hat{c}_{-p, i, \downarrow } \hat{c}_{p, i, \uparrow }  \rangle_i,
        \end{equation}
    where $N_s$ is the number of sites associated with momentum summation. We take the convention $g>0$ to represent the attractive interaction strength. We define the intercell disorder average $\langle \cdots \rangle_i \equiv 1/N \sum_i^N \langle \cdots \rangle$. We take the large-$N$ limit in the later calculations and fix $N_s=32$ when performing numerical simulations for the quench dynamics.

    The intracell SYK-type interaction reads
    \begin{equation}
	\begin{split}\label{eq:HI}
		\hat{H}_{\text{I}} =& \sum_{x, i<j;k<l\ } J^{(1)}_{x, ij;kl} \hat{c}^{\dagger}_{x,i,\uparrow} \hat{c}^{\dagger}_{x,j,\uparrow} \hat{c}_{x,k,\uparrow} \hat{c}_{x,l,\uparrow}  \\
		&+ \sum_{x, i<j;k<l}J^{(2)}_{x,ij;kl} \hat{c}^{\dagger}_{x,i,\downarrow} \hat{c}^{\dagger}_{x,j,\downarrow} \hat{c}_{x,k,\downarrow} \hat{c}_{x,l,\downarrow}, \\
	\end{split}
    \end{equation}
where the random couplings in $\hat{H}_{\text{I}}$ obey the following relations
	 \begin{equation}\label{eq:randomcoupling}
	 	\begin{split}
	 		\text{expectation}\ \ \ \ & \overline{ J^{(1)}_{x, ij;kl} }=\overline{ J^{(2)}_{x, ij;kl} }=0  \\
	 		\text{variance}\ \ \ \ & \overline{ J^{(1)}_{x, ij;kl} J^{(1)}_{x', ij;kl} }=\overline{ J^{(1)}_{x, ij;kl} J^{(1)}_{x', ij;kl} }=2 J^2/N \delta_{x x'}. \\
	 	\end{split}
	 \end{equation}
	 We have introduced two random couplings $J^{(1)}_{x, ij;kl}, J^{(2)}_{x, ij;kl}$, which correspond to uncorrelated random interactions for fermions with different spins. 
  
     There are two reasons for considering Eq.~\eqref{eq:H}. Firstly, it provides a concrete model for studying superconductors with strong interactions, allowing for investigations into both equilibrium properties and quantum dynamics within the large-$N$ limit. Secondly, it is known that the original SYK model exhibits non-Fermi liquid behavior \cite{StanfordRemarksSachdevYeKitaevModel2016}, with its low-energy manifold being dual to the Jackiw–Teitelboim gravity theory in 1+1-D \cite{YangConformalSymmetryIts2016}. By generalizing this model to higher dimensions, it becomes a valuable tool for understanding strongly correlated materials \cite{RevModPhys.94.035004}. Therefore, our model sheds light on the understanding of superconductivity in higher-dimensional non-Fermi liquids. Recently, several works have also proposed similar constructions for the superconductivity SYK model, albeit in different dimensionality or with correlated SYK interactions through Yukawa interaction with soft boson.\cite{KamenevSachdevYeKitaevSuperconductivityQuantum2020,FranzSuperconductingInstabilitiesSpinful2021,SchmalianQuantumCriticalEliashberg2022}. 
  
	\subsection{Methods for Thermal Equilibrium Properties}  
	Following the standard approach elaborated in related works \cite{gu2017local,ZhangDispersiveSachdevYeKitaevModel2017,ZhaiDisconnectingTraversableWormhole2021}, we define the retarded Green's function in terms of Nambu spinor representation $(\hat{\psi}_{p,j,1}, \hat{\psi}_{p,j,2}) = (\hat{c}_{p, j,\uparrow}, \hat{c}^{\dagger}_{-p, j,\downarrow})$
	\begin{equation}\label{eq:Green_func}
		\begin{split}
			&G^>_{j j';s s'}(p; t, t') \equiv -i \langle  \hat{\psi}_{p,j,s}(t) \hat{\psi}^{\dagger}_{p,j',s'}(t')  \rangle \delta_{j j'} \\
			&G^<_{j j';s s'}(p, t, t') \equiv i \langle \hat{\psi}^{\dagger}_{p,j',s'}(t') \hat{\psi}_{p,j,s}(t)  \rangle \delta_{j j'}, \\
		\end{split}
	\end{equation}
	where $s=1,2$ represents two components of the Nambu spinor. In the thermal equilibrium, all Green's functions are only functions of $t-t'$ due to the time-translational symmetry, with $G^{\gtrless}_{j j';s s'}(p; t, t') = G^{\gtrless}_{s s'}(p, t-t') \delta_{j j'}$. The diagonal of intracell index $j$ in Green's function is due to the disorder average, and later on, we will ignore the intracell index $j$ in Green's function for convenience. Furthermore, we introduce the retarded Green's function $G^{R/A}$ related to $G^{\gtrless}$ as 
	\begin{equation}\label{relationGRA}
		G^{R/A}_{s s'}(p; t, t')=\pm \Theta\left(\pm (t-t')  \right) \left(G^>_{s s'}(p;t,t')-G^<_{s s'}(p;t,t')\right),
	\end{equation}
	where $\Theta\left( t\right)$ is the Heaviside step function. By performing the Fourier transformation, Green's function can be represented on the momentum and frequency domain. 
	 \begin{equation*}\label{eq:fourier}
	 	G^{R/A}_{s s'}(p, \omega)= \int 
	 	 \diff t \ G^{R/A}_{s s'}(p, t) e^{- i \omega t}.
	 \end{equation*}
	 Then we can have the self-consistent Schwinger-Dyson equation for the retarded Green's function
	\begin{equation}\label{eq:S-D}
		(G^R)^{-1}(p, \omega) = ((G^0)^R)^{-1}(p, \omega) - \Sigma^R(\omega)
	\end{equation} 
    The bare Green's function corresponds to the BCS Hamiltonian 
    \begin{equation}\label{eq:Gbare}
        ((G^0)^R)^{-1}(p, \omega) = (\omega + i 0^+) \hat{\sigma}^0 - \epsilon(p) \hat{\sigma}^z + \Delta_i \hat{\sigma}^x.
    \end{equation}
    Here $\{\hat{\sigma}^0, \hat{\sigma}^r\}$, with $r=x,y,z$ are the Pauli matrix in the basis of Nambu spinor. In terms of Green's function, the equilibrium order parameter $\Delta_i$ could be separately written as
	\begin{equation}\label{eq:Delta_i}
		\Delta_i = 
		\begin{cases}
			\Delta_{i,0} \qquad (\text{constant}) & \text{Type \RN{1}}, \\
			-\sum_{p} i g_i G^<_{12}(p;t, t) /N_s& \text{Type \RN{2}}.\\
		\end{cases}
	\end{equation} 
    By taking the large-$N$ limit, and utilizing the tools of Keldysh contour  \cite{kamenev_2011}, the self-energy on the time domain could be written as
	\begin{equation} \label{eq:SelfE}
		\begin{aligned}
				\Sigma^{\gtrless}_{ss'}(t, t') &= 
			\begin{tikzpicture}[baseline={([yshift=-6pt]current bounding box.center)}, scale=1.1]
				\draw[thick,mid arrow] (-24pt,0pt) -- (-13pt,0pt);
				\draw[dashed,thick] (-13pt,0pt)..controls (-8pt,7pt) and (8pt,7pt)..(13pt,0pt);
				\draw[thick,mid arrow] (-13pt,0pt)..controls (-8pt,-13pt) and (8pt,-13pt)..(13pt,0pt);
				\draw[thick,mid arrow] (-13pt,0pt)..controls (-8pt,13pt) and (8pt,13pt)..(13pt,0pt);
				\draw[thick,mid arrow] (13pt,0pt) -- (-13pt,0pt);
				\draw[thick,mid arrow] (13pt,0pt) -- (24pt,0pt);
				
				\draw (-24pt,6pt) node{$x,t'$};
				\draw (24pt,6pt) node{$x,t$};
				\draw (-24pt,-6pt) node{$s'$};
				\draw (24pt,-6pt) node{$s$};
			\end{tikzpicture}\\
			&=  J^2 G^{\gtrless}_{ss}(x=0; t, t') G^{\lessgtr}_{ss}(x=0; t', t) G^{\gtrless}_{ss}(x=0; t, t') \delta_{s s'} \\
			&=  \frac{1}{N_s^3}\sum_{p_1,p_2,p_3} J^2 G^{\gtrless}_{ss}(p_1; t, t') G^{\lessgtr}_{ss}(p_2; t', t) G^{\gtrless}_{ss}(p_3; t, t') \delta_{s s'}, \\
		\end{aligned}
	\end{equation}
	where $G^{\gtrless}_{ss}(x; t, t')$ is the Fourier transformation of $G^{\gtrless}_{ss}(p; t, t')$ and the retarded/advanced self-energy are similarly defined as
	\begin{equation}\label{relationSigmaRA}
		\Sigma^{R/A}_{s s'}(t, t')=\pm \Theta\left(\pm (t-t')  \right) \left(\Sigma^>_{s s'}(t,t')-\Sigma^<_{s s'}(t,t')\right).
	\end{equation}
	 We notice the self-energy Eq.~\eqref{eq:SelfE} only has spin diagonal terms, since the coupling $J^{(1)}_{x, ij;kl}, J^{(2)}_{x, ij;kl}$ are not correlated. Besides, the $x=0$ in the Green's function results from the intercell disorder average in the Eq.~\eqref{eq:randomcoupling}.
	
	To solve the real-time Green's functions self-consistently, we introduce the spectral function as 
	\begin{equation}\label{eq:GR_rho}
		G^{R}_{s s'}(p,\omega) = \int \diff z  \frac{\rho_{s s'}(p, z)}{z - \omega + \iu 0^+}, 
	\end{equation}
	which implies $\rho_{s s'}(p, \omega)=-\text{Im}G^R_{s s'}(p, \omega)/\pi$. The greater and lesser Green's functions are associated with spectral function by the fluctuation-dissipation theorem as
	\begin{equation}
		\begin{aligned}
			G^<_{s s'}(p,\omega)&=2\pi i n_F(\omega)\rho(p, \omega)_{s s'},\\
			G^>_{s s'}(p,\omega)&=-2\pi i n_F(-\omega)\rho(p, \omega)_{s s'},
		\end{aligned}
	\end{equation}
	where $n_F(\omega)$ is the Fermi-Dirac distribution function.
	By using Eq.~\eqref{eq:S-D} and Eq.~\eqref{eq:SelfE}, one can iteratively obtain the equilibrium spectral functions and Green's functions.
	
\subsection{Methods for Non-equilibrium Dynamics} 

	To study the quench dynamics, we choose the real-time approach and utilize the Kadanoff-Baym equation on the Keldysh contour  \cite{kamenev_2011}, which describes the real-time evolution of $G^{\gtrless}$. Using Eq. \eqref{eq:SelfE} and applying the Langreth rules  \cite{stefanucci_van_Leeuwen_2013} on the Schwinger-Dyson equation, we find that  \cite{ZhaiDisconnectingTraversableWormhole2021}: 
	\begin{equation}
		\begin{aligned}\label{KBeq}
			i\partial_{t_1}&G^\gtrless(p; t_1,t_2)+ (-\epsilon(p)\hat{\sigma}^0+\Delta_f(t_1)\hat{\sigma}^x) G^\gtrless(p; t_1,t_2)=\\ &\int d t_3 (\Sigma^R(t_1,t_3)G^\gtrless(p; t_3,t_2)+\Sigma^\gtrless(t_1,t_3)G^A(p; t_3,t_2)), \\
			-i\partial_{t_2}&G^\gtrless(p; t_1,t_2)+G^\gtrless(p; t_1,t_2) (-\epsilon(p)\hat{\sigma}^0+\Delta_f(t_2)\hat{\sigma}^x)=\\ &\int d t_3 (G^R(p;t_1,t_3)\Sigma^\gtrless(t_3,t_2)+G^\gtrless(p;t_1,t_3)\Sigma^A(t_3,t_2)).
		\end{aligned}
	\end{equation}
	Similarly, we consider both the quantum proximity effect and self-consistent procedure of $\Delta(t)$. We summarize the two cases as 
	\begin{equation}\label{eq:Delta_f}
		\Delta_f(t>0) = 
		\begin{cases}
			 \Delta_{f,0} \qquad (\text{constant}) & \text{Type \RN{1}} ,\\
			 -\sum_{p} i g_f G^<_{12}(p;t, t)/N_s & \text{Type \RN{2}} .\\
		\end{cases}
	\end{equation}
	The quench protocol can be realized in the following manner. For $t_1,t_2<0 $, we require that $G^\gtrless(p; t_1,t_2)=G^\gtrless(p, t_{12})$. In other words, $G^\gtrless(p; t_1,t_2)$ is given by the equilibrium solution with initial order parameter $\Delta_i$ defined in the Eq.~\eqref{eq:Delta_i} correspondingly, which serves as the initial conditions for the real-time dynamics. For $t_1, t_2>0$, the system drives away from the equilibrium with the new order parameter $\Delta_f(t)$. Solving the differential equation of $G^\gtrless(t_1,t_2)$ with the Eq.~\eqref{eq:SelfE} and \eqref{KBeq} gives the quantum dynamics. We apply the second order Euler's method and choose the time domain cutoff to be $t / \Delta t \in  [-n_t, n_t]$ with $n_t = 2000$ and discrete time step $\Delta t = 20\beta/(n_t J)$. We have benchmarked the numerical error by testing the time translation invariance for Green's functions when we evolve the Green's functions without changing any parameters of the system.
 
	\begin{figure}[t]
		\centering	\includegraphics[width=0.55\linewidth]{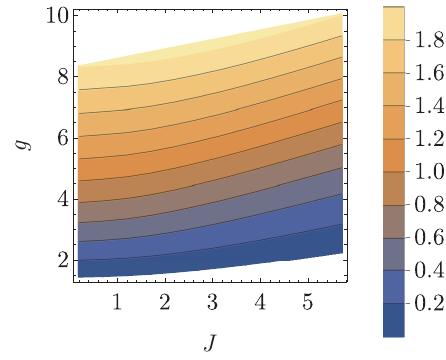}
		\caption{Critical BCS phase diagram with SYK type interaction. To illustrate the critical hyper-surface in the $g-T-J$ space, we project the hyper-surface into the $g-J$ plane using a contour plot. We fix the hopping strength $t_h=0.1$.}
		\label{fig:criticalphasediagram}
	\end{figure}

    \section{Numerical Results}
    In this section, we present numerical results both in thermal equilibrium and for quench dynamics. In both cases, we find the pairing strength is suppressed by the SYK random interactions. This qualitatively matches the observation in the cold atom experiment \cite{KohlDecayRevivalTransient2021}.

    \subsection{Phase Diagram in Thermal Equilibrium }  
    It is known that systems with attractive interactions exhibit superconducting transition at critical temperature $T_c$ for given BCS interaction strength $g$, or equivalently at critical BCS interaction strength $g_c$ at given temperature $T$ \cite{altland2010condensed}. In this part, we aim to explore the effect of SYK interaction $J$ on the transition temperature $T_c$ by computing the equilibrium phase diagram for type $\RN{2}$ models. The phase diagram for the traditional BCS system can be restored by taking $J\rightarrow 0$.
    
    The transition temperature $T_c$ can be determined by solving the gap equation with $\Delta=0$ \cite{altland2010condensed,zhai_2021}. Here we determine $T_c$ by taking a finite but small order parameter $\Delta=10^{-3}$ and perform the iteration for Green's functions for a fixed $J$ in the limit of $N_s\rightarrow \infty$ (see appendix \ref{suppsec:critical_param} for the details in taking the limit). After the Green's functions converge, $g$ is computed by using the relation \eqref{eq:Delta_i} (type $\RN{2}$). In numerics, we fix the hopping strength $t_h=0.1$. In fig.~\ref{fig:criticalphasediagram}, we show critical hypersurface in the $g-T-J$ space through a contour plot on the $g-J$ plane. We find with fixed $T$, larger $J$ leads to larger critical BCS interaction $g_c$. Since superconductivity occurs when the BCS interaction exceeds $g_c$, it suggests that the SYK interaction weakens superconductivity. This can be understood as the SYK interaction introduces a finite lifetime for fermions near the Fermi surface at a fixed temperature $T$. Consequently, it diminishes the pairing instability near the Fermi surface and leads to an increase in the critical BCS interaction $g_c$.
	
	We further compute the pairing strength for different SYK interaction strength $J$ with finite order parameter $\Delta$ or BCS interaction $g$ in type $\RN{1}$ and type $\RN{2}$ systems correspondingly. We define the pairing strength $\alpha \in (-0.5,0.5)$ as a 'normalized' order parameter, which reads
	\begin{equation}\label{eq:pairing}
		\alpha \equiv 1/N_s \sum_p \langle \hat{c}_{-p, i, \downarrow } \hat{c}_{p, i, \uparrow }  \rangle_i .
	\end{equation}  
	It corresponds to observing the magnetization in $x$ direct in the language of Anderson's pseudospin  \cite{AndersonRandomPhaseApproximationTheory1958}. This also indicates the close relation between the attenuating dynamics of fermionic superfluids and the magnetization dynamics of the random spin model \cite{ZhaiDisconnectingTraversableWormhole2021,ZhangOscillationHighT}
	
	\begin{figure}
		\centering
		\includegraphics[width=1.0\linewidth]{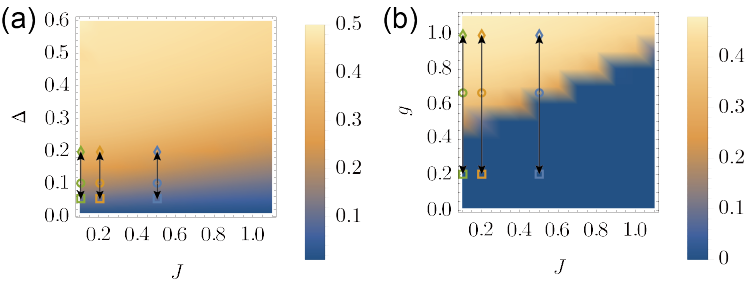}
		\centering
		\caption{The pairing $\alpha$ in the type $\RN{1}$ and $\RN{2}$ system. The color of the heatmap indicates the value of $\alpha$. We fix $t_h = 0.1$, $\beta=10$ and choose finite $N_s = 32$. (a) Type $\RN{1}$ system with proximity effect. The open markers label the initial order parameter $\Delta_i=0.1$ (circle) and quench to the final order parameter $\Delta_f = 0.1, 0.2$ (square and diamond). (b) Type $\RN{2}$ system with BCS iteration. The open markers label the initial BCS interaction $g_i=0.66$ (circle) and quench to the two sets of final BCS interactions $g_f = 0.2, 1.0$ (square and diamond). Green, orange, and blue colors indicate SYK interaction $J = 0.1, 0.2, 0.5$ separately.   }
		\label{fig:phaseDiagramPairing}
	\end{figure}
	
	The fig.\ref{fig:phaseDiagramPairing} (a), (b) show equilibrium pairing $\alpha$ in type $\RN{1}$ and type $\RN{2}$ system respectively. There are two remarks on the results. First, the proximity effect leads to a smooth change in the non-zero pairing $\alpha$ against the fixed order parameter $\Delta$. This can be understood as the magnetization induced by an external traverse magnetic field in the Anderson spin model, which is always non-zero. As a comparison, with BCS self-consistency, there is a typical second-order phase transition phenomenon at $g=g_c$  with each SYK interaction $J$. This is consistent with the original BCS theory \cite{zhai_2021,SchriefferTheorySuperconductivity1957}.
	Secondly, here we choose finite discretization of momentum $N_s=32$, for benchmarking the later calculation of quench dynamics. However, we find the $N_s=32$ result in fig.~\ref{fig:phaseDiagramPairing}(b) still qualitatively agree with the phase diagram illustrated in fig.~\ref{fig:criticalphasediagram} obtained in the limit of $N_s\to\infty$. Both of them show a positive correlation between critical $g$ and SYK interaction $J$. It provides a check for the validity of the finite $N_s$ calculations for the quench dynamics in the following subsection.

    \subsection{Attenuating Dynamics} 
	\begin{figure}
		\centering
        \includegraphics[width=0.9\linewidth]{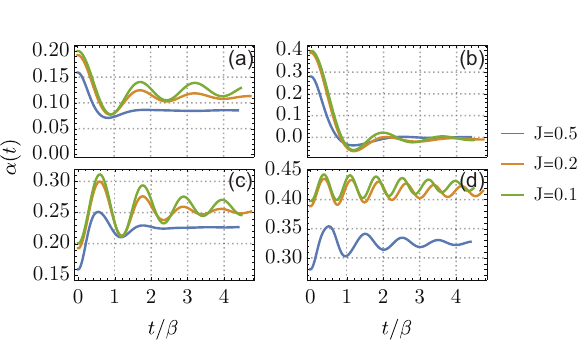}
		\caption{Quench dynamics of pairing in two types of system. We also fix $t_h = 0.1$, $\beta=10$ and choose finite $N_s = 32$ according to fig.~\ref{fig:phaseDiagramPairing}. (a, c) Quench dynamics with proximity effect. The system is initially prepared at $\Delta_i$ and quenched to $\Delta_f$. (a): $(\Delta_i, \Delta_f) = (0.1, 0.05)$; (c): $(\Delta_i, \Delta_f) = (0.1, 0.2)$. (b, d) Quench dynamics with order parameter self-consistency.  The interaction strength is initially $g_i$ and then quench to $g_f$. (b): $(g_i, g_f) = (0.66, 0.2)$; (d): $(g_i, g_f) = (0.66, 1.0)$. }
		\label{fig:quenchall}
	\end{figure}
	In the limit of $J\to 0$, our model is equivalent to the standard BCS mean-field theory. Using the time-dependent Bogoliubov theory, previous literature shows that small-amplitude oscillations of order parameter persist with a frequency of $2\Delta$, which is the energy of the Higgs mode\cite{volkovCollisionlessRelaxationEnergy,ZhangEvolutionHiggsMode2016a,KohlHiggsModeStrongly2018a,TsujiHiggsModeSuperconductors2020}. Later studies propose phase diagram with three dynamical phases classified according to the dynamics of the order parameter after a quantum quench. \cite{LevitovSynchronizationBCSPairing2006,DzeroDynamicalVanishingOrder2006,FosterQuantumQuenchPhase2015,GurarieNonequilibriumDynamicsWeakly2009}. The order parameter can disappear rapidly, damply oscillate, or persistently oscillate. However, these conclusions are obtained at the limit of large $N_s$ and assume a constant density of state. Here we can only take small $N_s$ and free fermion lattice dispersion in Eq.~\eqref{eq:Hbcs}, limited by the Green's function-based numerical method. We hereby focus on the dependence of the decaying rate on the SYK interaction parameter $J$.

 	In fig.~\ref{fig:phaseDiagramPairing}, we mark several open markers as the initial and final parameters for the quench dynamics numerics. We choose two sets of parameters: one set quenches to the system with a large order parameter (circle to diamond), and another one quenches to the system with a smaller order parameter(circle to square). In practice, it is realized by quenching superconducting order $\Delta$ induced by proximity effect in type $\RN{1}$ system, and by quenching BCS interaction $g$ in type $\RN{2}$ system. Besides, we are interested in the effect of SYK interaction $J$ on the non-equilibrium dynamics. Therefore, we also mark $J=0.1, 0.2, 0.5$ in the equilibrium phase diagram for comparison in the quench dynamics.
  
	The fig.~\ref{fig:quenchall} shows the oscillation is pervasive in different parameter regions. However, our numerical results show that the SYK-type interaction strongly attenuates the oscillation amplitude compared to the traditional BCS system. This qualitatively matches the absence of oscillation observed in the BCS-BEC quench experiment \cite{KohlDecayRevivalTransient2021}. Here we discuss the result of quench dynamics both with or without order parameter self-consistency. We fix hopping $t_h=0.1$ and inverse temperature $\beta=10$ but leave $J$ as an adjustable parameter. From fig.~\ref{fig:quenchall}, we simulate the quench dynamics with different $J$ and different iterative types. Fig.~\ref{fig:quenchall}(a), (c) belongs to the type \RN{1} system which quenches the background proximity order parameter, whereas Fig.~\ref{fig:quenchall}(b), (d) represents the type \RN{2} system which quenches the BCS interaction strength. For both types of systems, we find that when SYK-type interaction $J$ increases, the amplitude of the oscillation decreases. We recall that such decreasing in amplitude is consistent with the equilibrium phase diagram in fig.~\ref{fig:phaseDiagramPairing}, which indicates that the SYK interaction weakens the superconductivity.  

\begin{figure}
		\centering
		\includegraphics[width=0.8\linewidth]{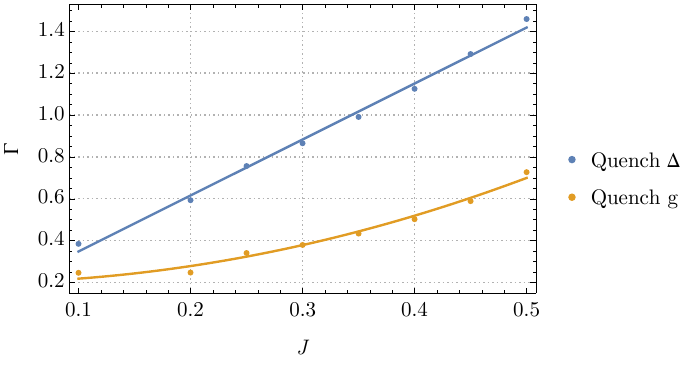}
		\caption{The decay rate in terms of different SYK interaction $J$. We also fix $t_h = 0.1$, $\beta=10$ and choose finite $N_s = 32$. The initial and final conditions correspond to fig.~\ref{fig:quenchall}(c) in the \textit{Quench $\Delta$} line, and correspond to fig.~\ref{fig:quenchall}(d) in the \textit{Quench $g$} line. }
		\label{fig:decayrate}
	\end{figure}

    It's worth exploring the decay rate against the SYK interaction. We fit the decay rate $\Gamma$ with formula $\alpha(t) \sim \alpha_0 e^{-\Gamma t} \cos(\Omega t + \theta) + c$, and the detail parameters are left to the appendix \ref{suppsec:decayrate}. As shown in fig.~\ref{fig:decayrate}, we find type \RN{1} and \RN{2} system exhibits different scaling laws for $J\gtrsim t_h$. The system with proximity effect shows perfect linear law, while the self-consistent BCS system shows quadratic scaling law. We argue this can be understood by a semi-classical Boltzmann equation in the limit of $t_h \rightarrow 0$\cite{zhai_2021}. We start with the type \RN{1} model describing the proximity effect. Given an order parameter $\Delta_f$, the system consists of Bogoliubov particles with energy $E_k=|\Delta_f|$. Without the SYK interaction $J$, the lifetime of the Bogoliubov particles is infinite. The quantum state after the quench can be viewed as a non-equilibrium state of Bogoliubov particles. The  relaxation of $\alpha$ is then because of the decay of Bogoliubov particles induced by the SYK interactions. Under the semi-classical approximation, this can be estimated by
    \begin{equation}
    \begin{aligned}
        \Gamma_k=&2\pi J^2\int \frac{dk_2dk_3dk_4}{(2\pi)^3} \delta(E_k+E_{k_2}-E_{k_3}-E_{k_4})\\
        &\times \left(n_F(E_{k_3})n_F(E_{k_4})(1-n_F(E_{k_2})) + (1-n_F(E_{k_3}))(1-n_F(E_{k_4}))n_F(E_{k_2})\right).
    \end{aligned}
    \end{equation}
    Here we focus on a two-to-two scattering for concreteness. Other scattering channels lead to a similar contribution. Unfortunately, the delta function diverges since $E_k=|\Delta_f|$. This divergence appears because we have assumed the quasi-particles have infinite lifetime: 
    \begin{equation}
        \delta(E_k+E_{k_2}-E_{k_3}-E_{k_4})=\int d\omega_2 d\omega_3 d\omega_4~\delta(E_k+\omega_{k_2}-\omega_{k_3}-\omega_{k_4})\prod_i\delta(\omega_i-E_{k_i}),
    \end{equation}
    where $\delta(\omega_i-E_{k_i})$ is the corresponding spectral function. When we take into account the finite lifetime of quasi-particles, the delta functions are smeared out and the divergence is cured. By dimensional counting, we expect the regularization $\delta(E_k+E_{k_2}-E_{k_3}-E_{k_4})\rightarrow 1/\Gamma$ for $t_h \rightarrow 0$. As a result, we find $\Gamma \sim J^2/\Gamma$, which indicates $\Gamma \propto J$. A similar phenomenon appears in the high-temperature limit of the Majorana SYK model \cite{zhang2021obstacle}. On the other hand, for the type \RN{2} model, the order parameter $\Delta$ is dynamical. As a result, the instantaneous spectral function of fermions in non-equilibrium dynamics is generally continuous in time. This indicates the lifetime of quasi-particles can be finite even without $J$. If this is the case, we expect the contribution from finite $J$ takes the form of $\Gamma\sim \Gamma_0+J^2/\Gamma_0$, which explains the quadratic dependence of $\Gamma$ with respect to $J$.

\vspace{0.05in}
\section{Discussion} 
In this work, we analyze the effect of SYK interactions in a one-dimensional BCS system. We employ the large-$N$ approximation and Green's function-based technique to solve the equilibrium problem and quench dynamics. Firstly, we calculate the critical hypersurface in the $g-T-J$ parameter space, which represents the superconductivity transition. Additionally, we investigate the equilibrium phase diagram of pairing with finite order parameters or BCS interaction. All these phase diagrams demonstrate that SYK interaction suppresses the superconductivity order. Using the phase diagram as a guide, we further explore the non-equilibrium quench dynamics. Our findings reveal that the oscillation of the pairing strength is damped by the SYK interaction $J$, which is consistent with the equilibrium phase diagram. This damping effect can be attributed to the introduction of interaction between Anderson spins, resulting in the thermalization of the Anderson spin system.

Our results represent the initial step towards achieving a comprehensive understanding of the attenuating dynamics observed in strongly interacting superconductors (or fermionic superfluids). For instance, it is reasonable to anticipate that the unitary Fermi gas would also undergo rapid thermalization due to the strong interactions between its constituent atoms \cite{KohlDecayRevivalTransient2021}. As a consequence, the relaxation of the pairing strength in this system should occur at a much faster rate compared to what is predicted by the traditional BCS theory. However, the development of a microscopic description for the quench dynamics in the unitary Fermi gas at low temperatures is deferred to future studies.

\section*{Acknowledgements}
We are especially grateful for the invaluable
discussions with Hui Zhai, whose advice is indispensable for
the whole work.

\bibliography{ref.bib}
\bibliographystyle{SciPost_bibstyle}
\nolinenumbers

\appendix
\section{Critical parameter obtained in $N_p\to\infty$ limit}
\label{suppsec:critical_param}

In this appendix, we show the technique to realize $N_p\to\infty$ in equilibrium calculation. We take care of the whole band dispersion and perform integration on the momentum space. 

Recalling the Schwinger-Dyson equation \eqref{eq:S-D}, \eqref{eq:Gbare}, now we will perform integration on the momentum index. 
    \begin{equation}
        G^R(x=0,\omega) = \int_{-\pi}^{\pi} \frac{\diff p}{2\pi} 
        \left(
            \begin{array}{cc}
             -\epsilon (p)-\Sigma_{11}^R(\omega )+\omega + \iu 0^+  & \Delta  \\
             \Delta  & \epsilon (p)-\Sigma_{11}^R(\omega )+\omega + \iu 0^+ \\
            \end{array}
            \right)^{-1}
    \end{equation}

Here we use the symmetry of Green's function, which is referred to previous work\cite{ZhaiDisconnectingTraversableWormhole2021,ZhangOscillationHighT}
\begin{equation}\label{eq:Sym1_GF}
\begin{split}
		G^>_{s_1 s_2}(t_1,t_2) 
		&= \begin{pmatrix}
			G^>_{2 2}(t_1,t_2) & G^>_{2 1}(t_1,t_2) \\
			G^>_{1 2}(t_1,t_2) & G^>_{1 1}(t_1,t_2) \\
		\end{pmatrix}_{s_1 s_2} \\
	\end{split}
\end{equation}
\begin{equation}\label{eq:Sym2_GF}
    \begin{split}
		G^>_{s_1 s_2}(t_1,t_2) 
		&= \begin{pmatrix}
			-G^<_{1 1}(t_2,t_1) & G^<_{1 2}(t_2,t_1) \\
			G^<_{1 2}(t_2,t_1) & -G^<_{1 1}(t_2,t_1) \\
		\end{pmatrix}_{s_1 s_2}. \\
    \end{split}
\end{equation}
Therefore we only need to consider the 11 and 12 components of the Green's function and self-energy. Also we remember $\Sigma^R_{12} = 0$ due to the format of SYK interaction Eq.~\eqref{eq:randomcoupling}.

Integrate over momentum leads to the final result
\begin{equation}
    \begin{split}
        G^R(x=0,\omega)_{11} &= -\int_{-\pi}^{\pi} \frac{\diff p}{2\pi} \frac{\epsilon (p) + \omega -\Sigma_{11}^R(\omega ) }{\epsilon (p)^2 + \Delta ^2 -\left(\omega -\Sigma_{11}^R(\omega )\right)^2} \\
        &= - \frac{\omega -\Sigma_{11}^R(\omega)}{\sqrt{A (A+4 t^2)}} (-1)^{\operatorname{Floor}\left[ \frac{\pi + \operatorname{Arg}[A+4t^2] - \operatorname{Arg}[A]}{2\pi} \right]  }, \\
    \end{split}
\end{equation}
and
\begin{equation}
    \begin{split}
        G^R(x=0,\omega)_{12} &= \int_{-\pi}^{\pi} \frac{\diff p}{2\pi} \frac{\Delta}{\epsilon (p)^2 + \Delta ^2 -\left(\omega -\Sigma_{11}^R(\omega )\right)^2} \\
        &= \frac{\Delta}{\sqrt{A (A+4 t^2)}} (-1)^{\operatorname{Floor}\left[ \frac{\pi + \operatorname{Arg}[A+4t^2] - \operatorname{Arg}[A]}{2\pi} \right]  }, \\
    \end{split}
\end{equation}
where the polynomial $A = \Delta ^2 - \left(\omega -\Sigma_{11}^R(\omega )\right)^2 $. We have already known that self-energy only depends on Green's function located at $x=0$. Hence, the integrated Schwinger-Dyson equation gives rise to a close form and can be solved self-consistently.

 \begin{figure}
	\centering
  \includegraphics[width=0.9\linewidth]{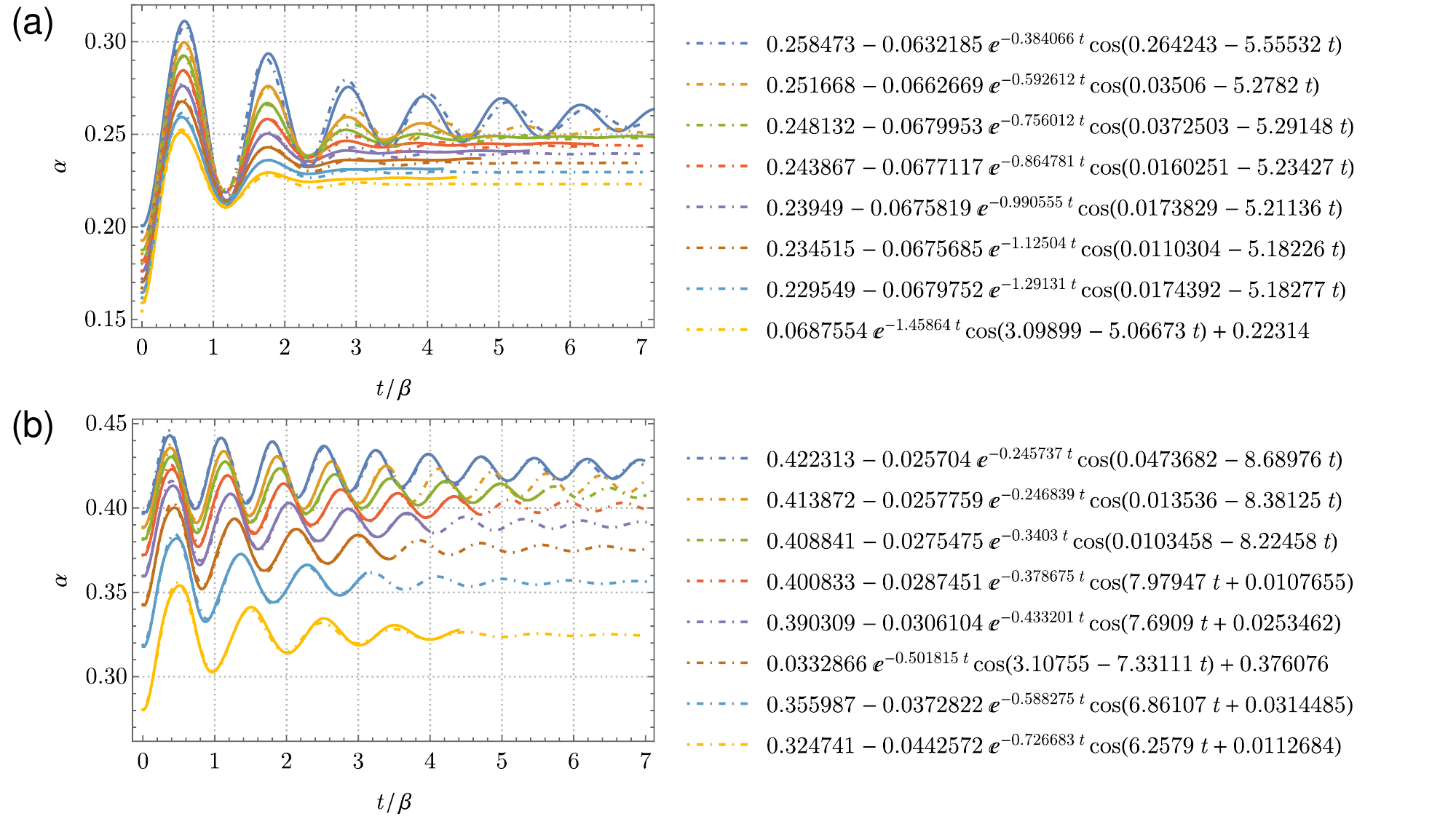}
		\caption{The fitting of decay rate in terms of different SYK interaction $J$. We fix $t_h = 0.1$, $\beta=10$ and choose finite $N_s = 32$. The initial and final conditions of (a) correspond to \textit{Quench $\Delta$} line in the fig.~\ref{fig:decayrate}, and (b) corresponds to the \textit{Quench $g$} line of fig.~\ref{fig:decayrate}. The right panel illustrates the detailed fitting formulas. From top to bottom, each curve corresponds to different SYK interactions $J=0.1, 0.2, 0.25, 0.3, 0.35, 0.4, 0.45, 0.5$.}
		\label{fig:suppFit}
	\end{figure}

\section{Details of fitting decay rate}
\label{suppsec:decayrate}

Different scaling laws of decay rate in quench $\Delta$ and quench $g$ protocols are revealed in fig.~\ref{fig:decayrate}. Here we fit  each $\alpha(t)$ curve with the first 280 data points for both quench protocols, as shown in fig.~\ref{fig:suppFit}. The fitting formula is
\begin{equation}
\Gamma = 
    \begin{cases}
        & 0.0799397 + 2.67588 J \qquad (\textrm{Quench $\Delta$}) \\
        & 0.197641 + 2.00575 J^2 \qquad (\textrm{Quench $g$}) \\
    \end{cases},
\end{equation}
which is obtained by Mathematica $\texttt{FindFit}$ formula.

For concreteness, we have tested the robustness of our conclusion by adjusting different fitting time periods in the data. There is no qualitative difference between different fitting regions.

\end{document}